\documentclass[%
  reprint,
  superscriptaddress,
  amsmath,amssymb,
  aps,
  prl,
  floatfix,
]{revtex4-2}

\usepackage{graphicx}
\usepackage{dcolumn}
\usepackage{bm}
\usepackage{hyperref}
\usepackage{xcolor}

\newcommand{\pourcentwidth}{0.99}                               
\newcommand{\nt}[1]           {#1}                              

\newcommand{\eps} {\varepsilon}

\newcommand{\op}[1] {\mathbf{#1}} 

\newcommand{\abs}[1]    { {\left\vert #1 \right\vert}}

\newcommand{\GF}[2] {\langle\!\langle\, {#1}; \, {#2}\,\rangle\!\rangle}

\begin{document}

\title{\nt{Quantum dot-based device for high-performance magnetic microscopy and spin filtering in the Kondo regime}}
\author{Pierre Lombardo}\email{pierre.lombardo@univ-amu.fr}
\affiliation{Aix-Marseille University, Faculty of Science, St J\'er\^ome, Marseille, France}
\affiliation{Institut Mat\'eriaux Micro\'electronique Nanosciences de Provence, UMR CNRS 7334, Marseille, France}
\author{Imam Makhfudz}
\affiliation{Aix-Marseille University, Faculty of Science, St J\'er\^ome, Marseille, France}
\affiliation{Institut Mat\'eriaux Micro\'electronique Nanosciences de Provence, UMR CNRS 7334, Marseille, France}
\author{Steffen Sch\"afer}
\affiliation{Aix-Marseille University, Faculty of Science, St J\'er\^ome, Marseille, France}
\affiliation{Institut Mat\'eriaux Micro\'electronique Nanosciences de Provence, UMR CNRS 7334, Marseille, France}
\author{Roland Hayn}
\affiliation{Aix-Marseille University, Faculty of Science, St J\'er\^ome, Marseille, France}
\affiliation{Institut Mat\'eriaux Micro\'electronique Nanosciences de Provence, UMR CNRS 7334, Marseille, France}

\date{\today}

\begin{abstract}
  We propose a nanoscale device consisting of a double quantum dot with a full exchange and pair hopping interaction.  In this design, the current can only flow through the upper dot, but is sensitive to the spin state of the lower dot. The system is immersed in a highly inhomogeneous magnetic field, and only the bottom dot feels a substantial magnetic field, while the top dot experiences only a residual one. 
 We show that our device exhibits very interesting magnetic field-dependent transport properties at low temperatures. The Kondo effect partially survives the presence of the magnetic field and allows  to obtain conductances that differ by several orders of magnitude for the two spin types across the top dot.
 Interestingly, as a function of the magnetic field, our two-dot device changes from a spin singlet state to a spin triplet state, in which the amplitudes of the spin-dependent conductances are reversed.
 Our device is able to discriminate between positive and negative magnetic fields with a high sensitivity and is therefore  particularly interesting for imaging the surface of anti-ferromagnetic (AF) insulating materials with alternated surface magnetic field, as well as for spin filtering applications.
  \end{abstract}

\maketitle


Magnetic structure of matter is a fascinating field, where a large variety of orders can be observed. In addition to ferromagnetic and antiferromagnetic orders, competing magnetic interactions give rise to very interesting complex non collinear spin states~\cite{C1987}.
Contrasting with ferromagnetic materials, in materials where antiferromagnetic interaction is dominant, it is often possible to switch magnetic moments at low energy cost. This can be promising for low-energy spintronic devices~\cite{JMW2016}. Spin filtering is another important ingredient of spintronics which may be realized with our proposal of capacitively coupled quantum dots. Spin filtering in transport processes via quantum dots had been proposed already several times~\cite{RSL2000,SSG2002,FHE2003,PB2006,SCA2006} but our proposal is interesting since one can reach very high polarizability as we will demonstrate in the following.

Nevertheless, investigating the magnetic properties of these materials is a difficult task because  local magnetic moment is either very small or oscillating at small length scales and they display insulating properties.
Indeed, magnetic force microscopy (MFM) is adapted to insulating materials  and can be used to unravel the magnetic properties of nanomaterials applied in biological systems and future directions for quantum technologies~\cite{WCA2023}. Unfortunately its sensitivity is not sufficient to detect very low or rapidly oscillating magnetic fields existing in antiferromagnets. 
Spin-polarized scanning tunneling microscopy is a much more sensitive technique and is a powerful technique to probe 
nano- and atomic-scale magnetism~\cite{WB2001}. However,   it is restricted to conductive materials.

Real space imaging using a single-spin magnetometer based on a single nitrogen–vacancy (NV) defect in diamond has been shown to be very efficient for imaging complex antiferromagnetic orders at the nanoscale~\cite{GAG2017}. 
The obtained image showed a spin cycloid in  a multiferroic bismuth ferrite (BiFeO$_3$ ) thin film with  a period of about 70 nanometers, consistent with values determined by macroscopic diffraction. However, a biased field has to be applied to determine the sign of the measured magnetic field. This  applied magnetic field could  influence  the fragile magnetic structure of the sample.

Here we propose a quantum dot-based device which is able to measure the local magnetization existing in the close proximity of an antiferromagnetic sample surface. Our device does not need any external applied magnetic field and takes advantage of the high sensibility of the Kondo effect.

Quantum dots can be used as highly sensitive magnetic field measurement. Recently, graphene quantum dots show promise as novel magnetic field sensors~\cite{GSP2023} by using confined massless Dirac fermions in electrostatically defined QDs.

Kondo effect plays a crucial role in transport properties of quantum dots. In the Coulomb blockade regime~\cite{AL1986,FD1987,LPWEUD1991,Kastner1992,Hewson1993book},  
the low temperature restoration of the conductance is one of the most spectacular manifestation of this effect.  The sharpness of the  zero-bias  resonance in the differential conductance~\cite{GSM1998,KSE2007,LSB2002,WFF2000,NCL2000,KMH2013,DRA2018},  opens up a wide range of highly sensitive applications using the Kondo effect
for the development of molecular electronics~\cite{MF2018,LTS2019}, for example.  

Like other quantum phenomena, the resonant spin-flip correlations underlying the Kondo effect are limited to very low temperatures and persist only up to some tens of milli-Kelvin in the usual semiconductor-based devices, or at best some tens of Kelvin in molecular electronics~\cite{MF2018,YN2004}, although Kondo temperatures as high as $100\,\mathrm{K}$ have been reported for magnetic impurities on surfaces~\cite{WDS2004,TGN2019}.

Besides its temperature dependance, the magnetic nature of the Kondo effect implies also a strong external magnetic field dependance. The interplay of Kondo effect and applied magnetic field has been studied in many previous works.
For example, nonequilibrium Kondo transport through a quantum dot in a magnetic field has been investigated in 2013~\cite{SG2013} and 
more recently, magnetic ﬁeld induced Kondo effect that takes place for level crossings in ﬁnite Co  chains has been studied by Danu et al.~\cite{DAM2019}
Temperature and magnetic field dependence of a Kondo system in the weak coupling regime has been investigated experimentally for  organic molecules weakly coupled to a Au (111) surface, showing a splitting of the resonance in magnetic field~\cite{ZKH2013}.
More recently, experimental results on a molecular quantum dot in the Kondo regime demonstrate that the bias-dependent thermocurrent is a sensitive probe of universal Kondo physics, directly measuring the splitting of the Kondo resonance in a magnetic field~\cite{HCV2022}.

\begin{figure}
\begin{center}
  \includegraphics[width=0.6\columnwidth]{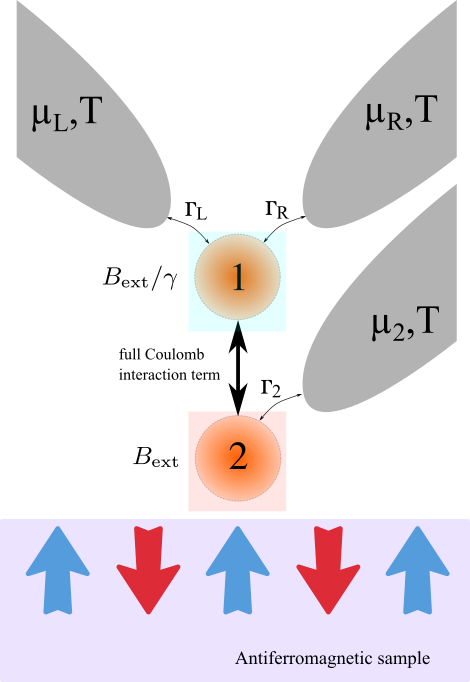}
  \caption{\label{fig0}
    Nanoscale \nt{device} where two single-level QDs, $\eps_1$ and $\eps_2$, with a full  Coulomb interaction term including exchange coupling and pair hopping, are connected to three uncorrelated metallic leads.   An electron can tunnel from the left (L) lead via  dot 1 to the right (R) lead, thus representing \nt{the channel of a transistor}.  The third lead controls the charge on  dot 2 which, although isolated from the channel, influences the current through the latter and thus acts as a gate. In the vicinity of an antiferromagnetic material, the magnetic field is very heterogeneous and dots 1 and 2 are respectively immersed in magnetic fields $B_{\mathrm{ext}}/\gamma$ and $B_{\mathrm{ext}}$.}
\end{center}
\end{figure}

The possibility of observing electron transport governed by the Kondo effect with an applied magnetic field is made possible here by the strong magnetic field gradient existing in the vicinity of the surface of the antiferromagnetic material. 
The device proposed in this Letter is schematically depicted in Fig.~\ref{fig0}. The conductance across  the first dot, labeled 1 (conduction channel) allows to measure the magnetic leakage field of the sample $B_{\mathrm{ext}}$. The second dot, labelled  2, is able to get very close to a magnetized sample  and is therefore immersed in $B_{\mathrm{ext}}$ while dot 1 experiences  $B_{\mathrm{ext}}/\gamma$.   The electronic  occupation of dot 2 is therefore expected to be strongly spin dependent, which ultimately pilots a spin polarized current across dot 1 via  interdot Coulomb interactions. The full Coulomb interaction term  contains exchange coupling and pair hopping processes and consequently correlates spin dependent properties of dot 1 and 2.  Experimental realization of capacitively coupled quantum dots without interdot net tunneling have already  been achieved e.g. in GaAs/AlGaAs heterostructures~\cite{CWM2002,HWDK2007}, and tunable interdot couplings with minimal residual interdot tunneling were implemented using bilayer graphene on silicon substrate~\cite{FVT2012}.

We will show in the following that the present  double-quantum dot setup not only offers the possibility to construct a quantum \nt{device} capable of measuring surface magnetic field but can also be used to generate an almost pure spin-current.

%


To investigate the setup proposed in Fig.~\ref{fig0}, we will study the corresponding single-impurity Anderson model~\cite{A1961} within the framework of the non-crossing approximation (NCA)~\cite{MJP2000,OK2006,ZCR2013}.  The model's electronic Hamiltonian contains contributions from the leads, the dots, and from tunneling between leads and dots, $H=H_\mathrm{leads}+H_\mathrm{dots}+H_\mathrm{tun}$.  The three uncorrelated metallic leads, $\alpha\in\{L,R,2\}$, are described by
  \begin{subequations}
    \label{eq:Ham}
    \begin{equation}
      H_{\mathrm{leads}}\ =\ \sum_{\alpha k \sigma} \eps_{\alpha k}\,
      \op{c}_{\alpha k\sigma}^{\dagger}\, \op{c}_{\alpha k\sigma}  \,
      \text{,}
      \label{eq:Hleads}
    \end{equation}
    with $\op{c}_{\alpha k\sigma}^{\dagger}$ and $\op{c}_{\alpha k\sigma}$ the creation and annihilation operators of $\sigma$-electrons in state $k$ on the lead $\alpha$.  The contribution from the dots is~\cite{YWD2011}

\begin{eqnarray}
&H_{\mathrm{dots}}& =
\,  \sum_{i \in \{1,2\},\sigma} \eps_{i\sigma}\,\op{n}_{i \sigma}
\,+\, \sum_{i \in \{1,2\}}U_i \op{n}_{i\uparrow}\op{n}_{i\downarrow}-2J \hat{\bf{S}}_{1}\hat{\bf{S}}_{2} \nonumber \\
 &+&  (U^{\prime}-\frac{J}{2})\sum_{\sigma,\sigma^{\prime}}\op{n}_{1\sigma}\op{n}_{2\sigma^{\prime}} 
 \nonumber \\
&+&  J^{\prime}\sum_{i \neq i^\prime \in \{1,2\}} \op{d}_{i\uparrow}^{\dagger}\op{d}_{i\downarrow}^{\dagger}\op{d}_{i^\prime\downarrow}\op{d}_{i^\prime\uparrow}
 \label{eq:Hdots}
\end{eqnarray}

    where $\op{n}_{i}=\op{n}_{i\uparrow}+\op{n}_{i\downarrow}$ is the total electronic occupation on dot $i$, with $\op{n}_{i\sigma}=\op{d}_{i\sigma}^{\dagger}\op{d}_{i\sigma}$ the number of $\sigma$-electrons on the dot, and $\op{d}_{i\sigma}^{\dagger}$ ($\op{d}_{i\sigma}$) the corresponding creation (annihilation) operators. The sum in the above expression thus describes two individual Anderson dots, with the  dot 2 serving as a magnetic gate,  and the dot 1 defining the conduction channel. Eq.~(\ref{eq:Hdots}) accounts for the full interaction term between both dots and can be implemented experimentally as described in Refs.~\onlinecite{CWM2002,HWDK2007,FVT2012}.
$\varepsilon_{1\sigma}$ and  $\varepsilon_{2\sigma}$ are the spin dependent  energy levels of the  dots, including Zeeman splittings due to external magnetic fields $B_{\mathrm{ext}}/\gamma$ for dot 1 and $B_{\mathrm{ext}}$ for dot 2. We have $\varepsilon_{1\sigma}=\varepsilon_{1}-\sigma B_{\mathrm{ext}}/\gamma$ and $\varepsilon_{2\sigma}=\varepsilon_{2}-\sigma B_{\mathrm{ext}}$. In the following we take $\gamma=10$ to take account of the inhomogeneity of the magnetic field at the surface of the antiferromagnetic material,
 and $\eps_1=\eps_2$ to reduce the number of independent parameters, but it would be entirely possible to control these two energy levels independently.  
The operator $\hat{\bf S}_{i}= (\hat S^{x}_{i},\hat S^{y}_{i},\hat S^{z}_{i})$ 
is the $1/2$-spin operator for dot $i$. 
The parameters $U_i$ denote the repulsions between electrons occupying the same dot $i$, $U^{\prime}$ is the repulsion between electrons
occupying different dots, $J$ is the Hund exchange term, and $J^{\prime}$ is the double hopping term. This last term is systematically present when exchange interactions take place. 
Note that the Hamiltonian $H_{\mathrm{dots}}$ is  relevant for double-quantum dots and for single quantum dot with two orbitals, however there is no  reason to enforce  invariance  under rotation in the orbital space for the double-quantum dots, and the constraints $U^{\prime}=U-J-J^\prime$ and $U_1=U_2=U$ are not required.

    Finally, the lead-dot tunneling is given by
    \begin{multline}
      H_{\mathrm{tun}}\, =\, \sum_{\substack{\alpha\in\{L,R\}\\k\sigma}}
      \left( t_{\alpha k}\, \op{d}^{\dagger}_{1\sigma}\,\op{c}_{\alpha k\sigma}
      \,+\, t^*_{\alpha k}\, \op{c}^{\dagger}_{\alpha k\sigma}\,\op{d}_{1\sigma} \right)
      \\
      + \sum_{k\sigma}
      \left( t_{2 k}\, \op{d}^{\dagger}_{2\sigma}\,\op{c}_{2 k\sigma}
      \,+\, t^*_{2 k}\, \op{c}^{\dagger}_{2 k\sigma}\,\op{d}_{2\sigma} \right)
      \,\text{,}
      \label{eq:Htun}
    \end{multline}
    where the first line describes the coupling of electrons on dot 1 to the left and right lead, while the second line allows electrons on dot 2 to tunnel to the third lead.  
  \end{subequations}
  For most practical cases, the tunneling amplitudes $t_{\alpha k}$ only depend on $k$ via the corresponding level energy, $\eps_{\alpha k}$, such that the tunneling to each lead is entirely summarized by the hybridization strength~\cite{JauhoWingreenMeir1994} $\Gamma_{\alpha}(\eps)= 2\pi \abs{t_\alpha(\eps)}^2 N_{\alpha}(\eps)$ where $N_{\alpha}(\eps)$ is the spin-summed DOS on lead $\alpha$ ($\in\{L,R,2\}$).
  Here, we further assume that all three leads are identical metals, with a DOS characterized by a single wide band, and that the tunneling between any lead and its dot is governed by the same only weakly energy-dependent physical process such that all three hybridization strengths are the same, $\Gamma_{L}(\eps)=\Gamma_{R}(\eps)=\Gamma_{2}(\eps)$.  Specifically, we take broad Gaussian lead DOSes, of half-width $D=70\,\Gamma$, where $\Gamma=\Gamma_L(\bar{\mu})+\Gamma_R(\bar{\mu})$ is the total hybridization strength of the lower dot at the mean chemical potential $\bar{\mu}=\frac{1}{2}\left(\mu_L+\mu_R\right)$. $\Gamma$ will henceforth serve as our energy unit \nt{which, with $k_{\mathrm{B}}$ set to unity and the Bohr magneton $\mu_{\mathrm{B}}$ set to 2, will also be our unit of temperature and magnetic field.}   Furthermore, we assume zero voltage and temperature bias such that $\bar{\mu}=\mu_L=\mu_R=\mu_2$ and $T=T_L=T_R=T_2$.


  The spin dependent  conductance $G^\sigma(T) = e^2 I_0^\sigma(T)$ is given in terms of spin resolved energy-weighted integrals~\cite{KH2003},
  \begin{equation}
    I_n^\sigma(T)= \frac{2}{h}\int  \eps^n   \left( -\frac{\partial f_\sigma}{\partial \eps} \right) \tau_\sigma^{\rm eq}(\eps)     {\rm d}\eps\;\text{.}
  \end{equation}
  The integrand comprises the spin dependent Fermi function, $f_\sigma(\eps)$, and the equilibrium transfer function, $\tau_\sigma^{\rm eq}(\eps)=\frac{\pi}{4} A^\sigma_1(\eps)\Gamma(\eps)$.  Similarly, the occupancies of the dots $j=1,2$ also follow from the lead Fermi functions and the dot spectral functions $A^\sigma_j(\eps)$ via~\cite{WJM1993}
  \begin{displaymath}
    n^\sigma_j=\int f_\sigma(\eps-\mu_{j}) A^\sigma_j(\eps){\rm d}\eps\,\text{.}  
  \end{displaymath}

The $A^\sigma_j(\eps)=-\frac{1}{\pi}{\rm Im}G^{\rm r}_{j\sigma}(\eps+i\delta)$  are readily obtained from the corresponding retarded dot Green's functions $G^{\rm r}_{j\sigma}(\eps+i\delta)=\GF{\op{d}_{j\sigma}}{\op{d}_{j\sigma}^{\dagger}}$.  In the atomic limit, the dot Green's functions would show sharp transitions between the 4 local eigenstates of each dot.  The hybridization with the leads and the interdot Coulomb repulsion, however, create correlations and fluctuations between these 16 local states, and we address the latter within the framework of the NCA, an approximation suitable for resonant spin-fluctuations underlying the Kondo effect which are expected in the parameter regime under investigation.
Energies of the 16 local states are displayed in table~\ref{tab1}.
\begin{table}[htb]
\begin{center}
\begin{tabular}{|c|c|c|c|}
\hline
State & Occupation & Energy$-\varepsilon_0$   \\ 
\hline
$|\alpha_1\rangle=|0,0\rangle$ & 0 & 0   \\ 
\hline 
\hline 
$|\alpha_2\rangle=|\uparrow,0\rangle$ & 1  & $\varepsilon_1-B_{\mathrm{ext}}/\gamma$ \\ 
\hline
$|\alpha_3\rangle=|\downarrow,0\rangle$ & 1  & $\varepsilon_1+B_{\mathrm{ext}}/\gamma$  \\ 
\hline
$|\alpha_4\rangle=|0,\uparrow\rangle$ & 1  & $\varepsilon_2-B_{\mathrm{ext}}$  \\ 
\hline
$|\alpha_5\rangle=|0,\downarrow\rangle$ & 1  & $\varepsilon_2+B_{\mathrm{ext}}$  \\ 
\hline
\hline
$|\alpha_6\rangle=|0,\uparrow\downarrow\rangle$ & 2  & Not an eigenvector  \\ 
\hline
$|\alpha_7\rangle=|\uparrow\downarrow, 0\rangle$ & 2  & Not an eigenvector  \\ 
\hline
$|\alpha_8\rangle=|\uparrow,\downarrow\rangle$ & 2  & Not an eigenvector  \\ 
\hline
$|\alpha_9\rangle=|\downarrow,\uparrow\rangle$ & 2  & Not an eigenvector  \\ 
\hline
$|\alpha_{10}\rangle=|\uparrow,\uparrow\rangle$ & 2  & $\varepsilon_1+\varepsilon_2+U^\prime-J-B_{\mathrm{ext}}-B_{\mathrm{ext}}/\gamma$  \\ 
\hline
$|\alpha_{11}\rangle=|\downarrow,\downarrow\rangle$ & 2  & $\varepsilon_1+\varepsilon_2+U^\prime-J+B_{\mathrm{ext}}+B_{\mathrm{ext}}/\gamma$  \\ 
\hline
\hline
$|\alpha_{12}\rangle=|\uparrow,\uparrow\downarrow\rangle$ & 3  & $\varepsilon_1+2\varepsilon_2 +U_2+2U^\prime-J-B_{\mathrm{ext}}/\gamma$ \\ 
\hline
$|\alpha_{13}\rangle=|\downarrow,\uparrow\downarrow\rangle$ & 3  & $\varepsilon_1+2\varepsilon_2+U_2+2U^\prime-J+B_{\mathrm{ext}}/\gamma$  \\ 
\hline
$|\alpha_{14}\rangle=|\uparrow\downarrow,\uparrow\rangle$ & 3  & $2\varepsilon_1+\varepsilon_2+U_1+2U^\prime-J-B_{\mathrm{ext}}$  \\ 
\hline
$|\alpha_{15}\rangle=|\uparrow\downarrow,\downarrow\rangle$ & 3  & $2\varepsilon_1+\varepsilon_2+U_1+2U^\prime-J+B_{\mathrm{ext}}$  \\ 
\hline
\hline
$|\alpha_{16}\rangle=|\uparrow\downarrow,\uparrow\downarrow\rangle$ & 4  & 
$2\varepsilon_1+2\varepsilon_2+U_1+U_2+4U^\prime-2J$  \\ 
\hline
\end{tabular} 
\caption{\label{tab1}Local energies of  $H_{\mathrm{dots}}$. 
Exchange and pair-hopping terms in $H_{\mathrm{dots}}$ impose the diagonalization of states $|\alpha_6\rangle$ to $|\alpha_9\rangle$.
}
\end{center}
\vspace{-0.6cm} 
\end{table}

In the subspace $\{|\alpha_6\rangle,|\alpha_7\rangle\}$, 
$$
H_{\mathrm{dots}}=\begin{pmatrix}2\varepsilon_1+U_1&J^\prime\\
				J^\prime&2\varepsilon_2+U_2 \end{pmatrix}
$$
 can be diagonalized using eigenvectors $|6\rangle$ and $|7\rangle$ which are simple linear combinations of $|\alpha_6\rangle$ and $|\alpha_7\rangle$.

In the subspace $\{|\alpha_8\rangle,|\alpha_9\rangle\}$, 
$$
H_{\mathrm{dots}}=\begin{pmatrix}
 H_{\mathrm{dots}}[8,8] &  -J\\
	-J  &   H_{\mathrm{dots}}[9,9]
	\end{pmatrix}
$$
where $H_{\mathrm{dots}}[8,8]=\varepsilon_1+\varepsilon_2+U^\prime+B_{\mathrm{ext}}-B_{\mathrm{ext}}/\gamma$ and $H_{\mathrm{dots}}[9,9]=\varepsilon_1+\varepsilon_2+U^\prime-B_{\mathrm{ext}}+B_{\mathrm{ext}}/\gamma$. It can be diagonalized using eigenvectors $|8\rangle$ and $|9\rangle$ which are simple linear combinations of $|\alpha_8\rangle$ and $|\alpha_9\rangle$.

\begin{figure}
\begin{center}
  \includegraphics[width=0.99\columnwidth]{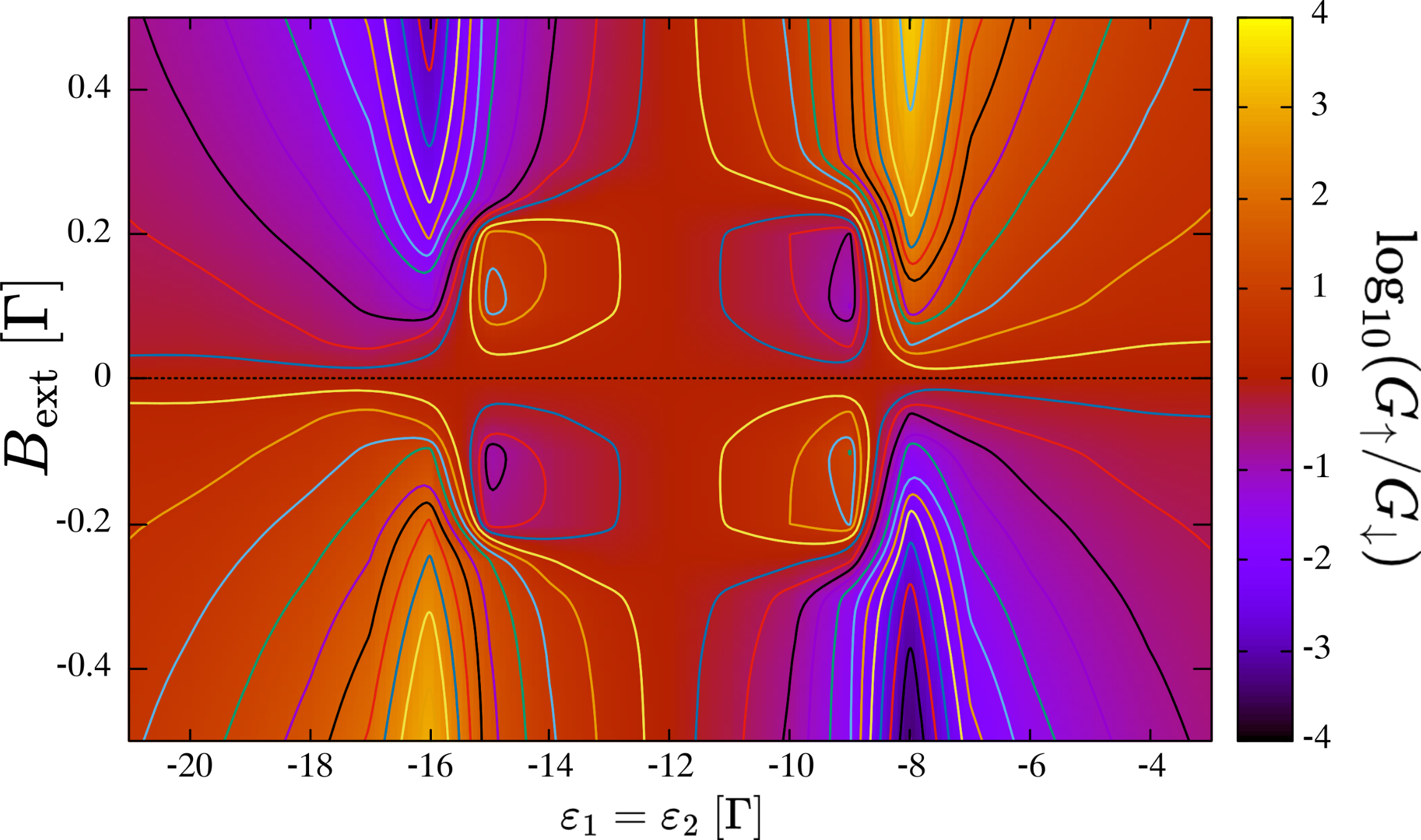}
  \caption{
  $\log_{10} (G_\uparrow/G_\downarrow)$ with respect to $\varepsilon_1=\varepsilon_2$ and $B_{\mathrm{ext}}$, for low temperature $T=5\times 10^{-4}$ . Parameters are  $J=J^\prime=0.2$, and $U_1=U_2=U^\prime=8$.
    \label{diagramme_de_phase_BT}}
\end{center}
\end{figure}
To give an overview of the possibilities of the device we present here, we show in Fig.~\ref{diagramme_de_phase_BT} a diagram in which the logarithm of the ratio of spin-polarized conductances $G_\uparrow/G_\downarrow$ is plotted as a function of local energies $\varepsilon_1=\varepsilon_2$ and magnetic field $B_{\mathrm{ext}}$. The temperature $T=5\times 10^{-4}$ is low enough for the Kondo effect to play a crucial role in the transport properties, as will be explained later in this article. All parameters are in $\Gamma$ units. Coulomb interaction parameters are $J=J^\prime=0.2$, and $U_1=U_2=U^\prime=8$. In the diagram, $\varepsilon_1=\varepsilon_2$ goes from -21 to -3 and $B_{\mathrm{ext}}$ from -0.5 to 0.5. The $\gamma$ parameter representing the inhomogeneity of the magnetic field at the sample surface is 10. 
The diagram shows four well-defined zones in which conductance is strongly spin-polarized. This polarization occurs when a magnetic field is applied to the device. Remarkably, the sign of $\log_{10} (G_\uparrow/G_\downarrow)$ gives direct information on the sign of $B_{\mathrm{ext}}$, suggesting applications for direct imaging of antiferromagnetic surfaces. Indeed, alternative techniques such as single-spin magnetometer based on a single nitrogen–vacancy defect in diamond~\cite{GAG2017} require an additional external field to be applied in order to discriminate the sign of the measured field. Such a procedure, which could affect the measurement of the magnetic field emanating from the sample, is not necessary with the device we propose.
Another important feature of the diagram in the figure~\ref{diagramme_de_phase_BT} is the extremely high degree of spin polarization obtained. For well-tuned parameters, $G_\uparrow/G_\downarrow$ can reach $10^{3}$. This property makes our device highly sensitive to the magnetic field emanating from the sample, but could also enable us to generate extremely pure spin-polarized currents. What is even more interesting is that the device can be switched from spin up filter to spin down filter very quickly by simply controlling the  gate voltage $\varepsilon_1=\varepsilon_2$, which could be very useful in spintronics for spin filtering.

\begin{figure}
\begin{center}
  \includegraphics[width=\pourcentwidth\columnwidth]{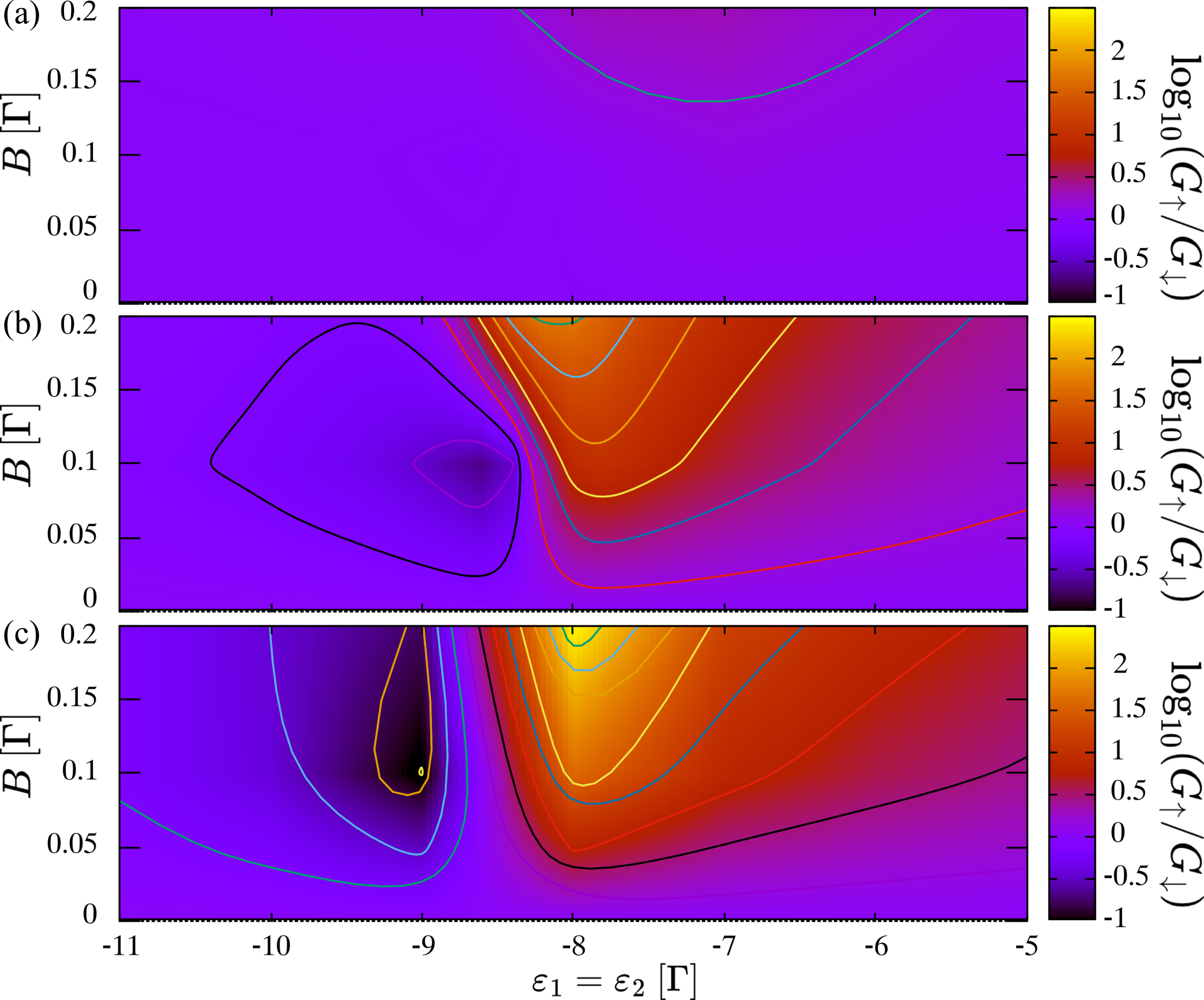}
  \caption{
  $\log_{10} (G_\uparrow/G_\downarrow)$ with respect to $\varepsilon_1=\varepsilon_2$ and $B_{\mathrm{ext}}$, for different temperatures (a) $T=5\times 10^{-2}$, (b) $T=5\times 10^{-3}$ and (c) $T=5\times 10^{-4}$. Parameters are  $J=J^\prime=0.2$, and $U_1=U_2=U^\prime=8$.
    \label{figure_zoomed}}
\end{center}
\end{figure}
In the following, we will see that the high sensitivity of the device is due to the Kondo effect, which induces a strong magnetic coupling between the spin states of dots 1 and 2. The existence of the Kondo effect under an applied magnetic field has been studied in many works, and it is well known that the magnetic field destroys the Kondo resonance through Zeeman splitting. In our device, this is not the case, as only dot 2 is immersed in a strong magnetic field, while dot 1 is exposed to only a weak magnetic field, allowing the Kondo effect to survive.
In fact, as we will see below, the Kondo effect only persists for a unique spin species on the channel dot, implying a strong asymmetry between the conductances for each spin species.

Transport through dot 1 is then highly affected by this magnetic effect, strongly differentiating the two electron spin states. Indeed, a first indication of the Kondo nature of transport is given in figure~\ref{figure_zoomed}, where the ratio of spin-resolved conductances is plotted for different temperatures. In part (a) of the figure, the temperature is $T=5\times 10^{-2}$ and even under optimum conditions, for $B_{\mathrm{ext}}=0.5$ (not shown), $\log_{10} (G_\uparrow/G_\downarrow)$ only reaches the value 0.15, i.e. $G_\uparrow/G_\downarrow\approx 1.4$. The device is very insensitive.  Sensitivity increases sharply as the temperature is lowered, and for $T=5\times 10^{-4}$, significant values of $G_\uparrow/G_\downarrow\approx 10^{3}$ are reached at moderate field strength  $B_{\mathrm{ext}}=0.2$.

\begin{figure}
\begin{center}
  \includegraphics[width=\pourcentwidth\columnwidth]{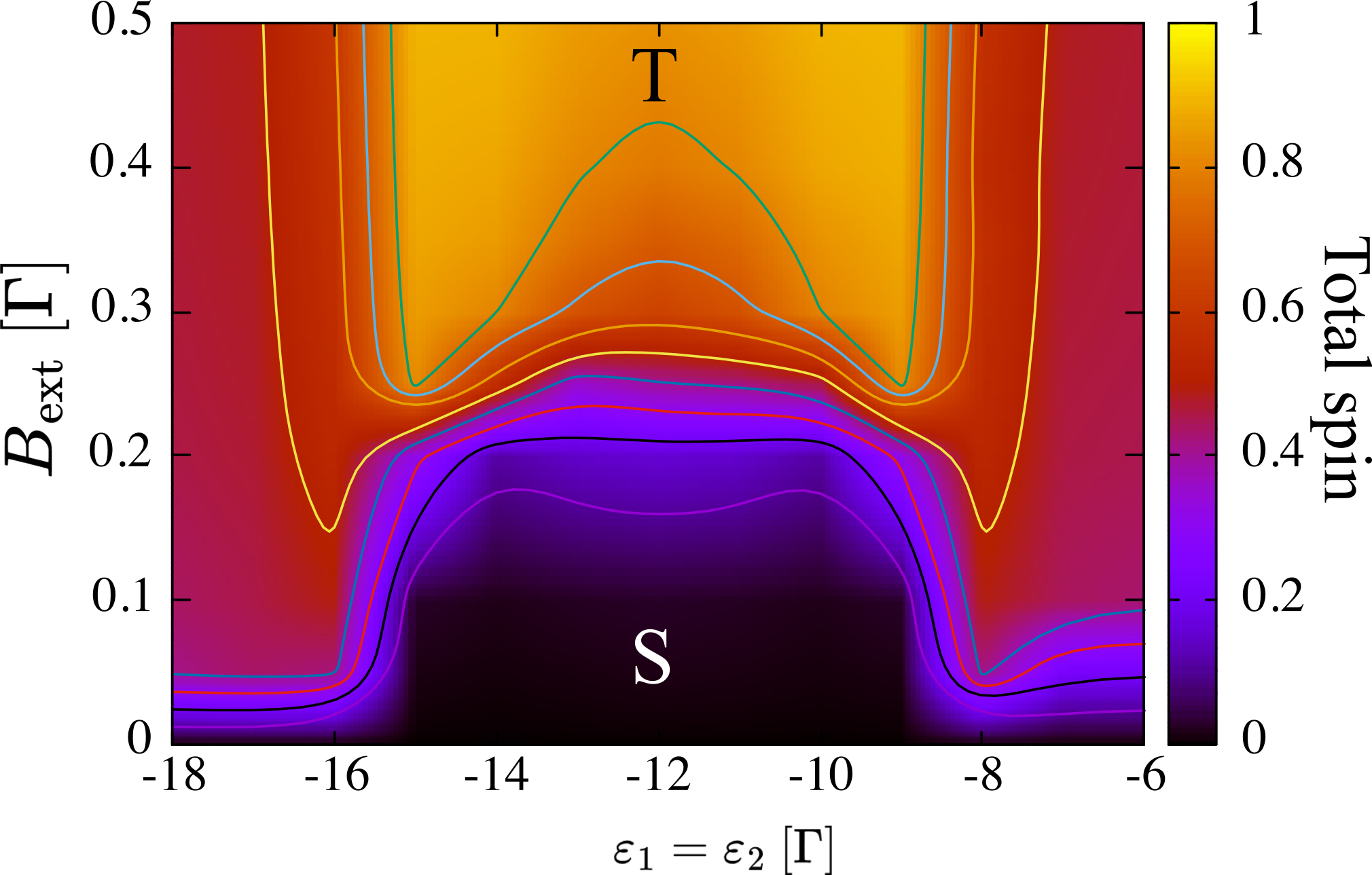}
  \caption{
  Total spin of the two-dot device $(n_1^\uparrow-n_1^\downarrow)/2+(n_2^\uparrow-n_2^\downarrow)/2$ with respect to 
  $\varepsilon_1=\varepsilon_2$ and $B_{\mathrm{ext}}$, for temperature $T=5\times 10^{-4}$. Parameters are  $J=J^\prime=0.2$, and $U_1=U_2=U^\prime=8$.
    \label{fig_ST}
    }
\end{center}
\end{figure}
In order to better understand how the spin states of two dots can be coupled thanks to the exchange term of the Coulomb interaction between dots, we show in Fig.~\ref{fig_ST} how the total spin of the two-dot device evolves.  Under magnetic field $B_{\mathrm{ext}}$, dot 2 gets almost entirely polarized, parallel to the applied magnetic field.
For site energies between -15 and -9, the polarization of dot 1 behaves differently depending on the value of the magnetic field.  At high magnetic fields (above 0.2), dot 1 polarizes parallel to the applied field (as with dot 2), and the resulting total spin is 1 (zone {\bf T} in Fig.~\ref{fig_ST}). This two-dot state has strong similarities with a triplet spin state. The singlet state is obtained at lower magnetic fields (below 0.2). In this case, dot 1 polarizes antiparallel to the applied magnetic field. The spin of dot 1 is governed by antiferromagnetic inter-dot exchange coupling and Kondo physics. The resulting total spin is 0 (zone {\bf S} in Fig.~\ref{fig_ST}).
\begin{figure}
\begin{center}
  \includegraphics[width=\pourcentwidth\columnwidth]{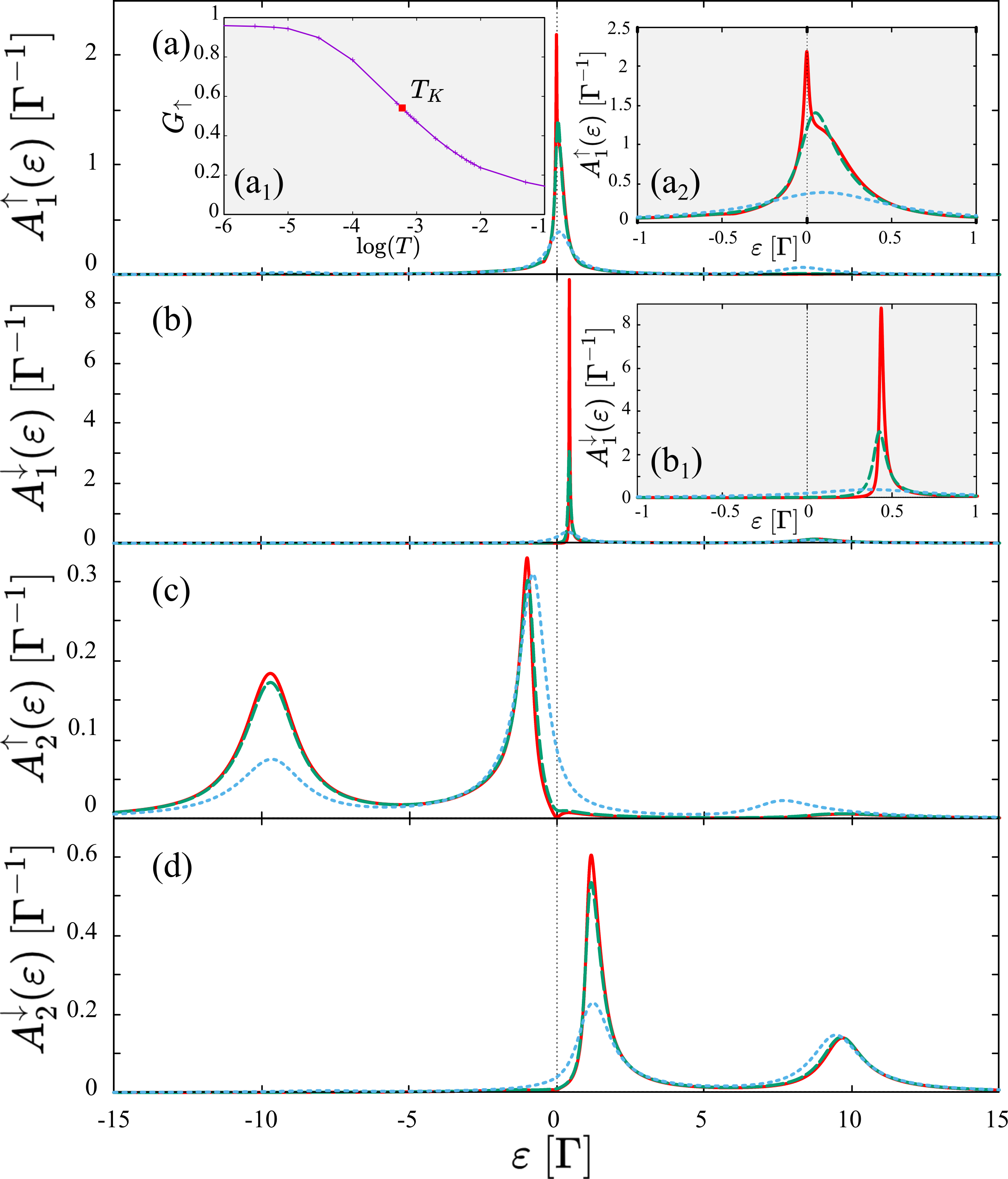}
  \caption{
   Dot spectral functions ((a),(b) for dot 1 and (c),(d) for dot 2) for $J=J^\prime=0.2$, $\varepsilon_1=\varepsilon_2=-8$, $T=0.5$ (blue-dotted line), $T=5\times 10^{-2}$ (green-dashed line) and $T=5\times 10^{-3}$ (red-solid line),  $B=1$. $U_1=U_2=U^\prime=8$. Inset (a$_1$) represents spin up polarized conductance as a function of temperature. Insets (a$_2$) and (b$_1$) show the low-energy behavior of spectral functions. 
    \label{DOS_plusieurs_Tre}}
\end{center}
\end{figure}
Finally, to gain a deeper understanding of how the Kondo effect impacts spin-resolved transport properties, we have plotted in Fig.~\ref{DOS_plusieurs_Tre} the $A_j^\sigma$ spectral functions for each dot and spin, for the three temperatures $T=0.5$, $T=5\times 10^{-2}$ and $T=5\times 10^{-3}$. 
These results therefore give an indication of how the Kondo effect takes hold in the device.
Indeed, inset (a$_1$) shows the evolution of spin-up conductance as a function of temperature, enabling us to define the Kondo temperature for the chosen physical parameters $T_{\mathrm{K}}=5\times 10^{-4}$.
For dot 1, parts (a) and (b) of the figure show the spin-up and spin-down spectral functions for these three temperatures, respectively. Parts (c) and (d) relate to dot 2. 
Under a positive magnetic field (here $B_{\mathrm{ext}}=1$), the dot 2 spectral functions are strongly polarized by Zeeman splitting at all three temperatures explored. Temperature has only a slight effect at high energies, but greatly reduces the density of low-energy excitations, close to $\varepsilon=0$. 
In dot 1, the effect of temperature is much more pronounced, leading to strong discrimination between spins. The spectral function of the down spin is significantly lowered near $\varepsilon=0$, while that of the up spin is subject to a pronounced Kondo resonance at low temperature. The effect of this resonance is to boost the conductance of the up spin, which explains the significant increase in low-temperature sensitivity of the device described in this paper.

\section{Conclusion}

In summary, we examined the electronic transport through a T-shaped double quantum dot with strong intra- and interdot Coulomb repulsions including exchange coupling and pair-hopping processes.
In this setup, electrons may only flow from source to drain via the upper dot.  The role of the lower dot is to monitor the upper dot spin-flow via an interdot interaction, depending on magnetic field on the lower dot. 
We have shown that the exchange interaction between dots enables the establishment of a Kondo regime at low temperatures. The device is then able to discriminate very selectively between the two spin states as a function of the applied magnetic field. 
We have shown that the selective development of an Abrikosov-Suhl-Kondo resonance for a unique spin species on the conduction dot makes it possible to achieve very large values of the spin conductance ratio.
Potential applications in magnetic microscopy and spin filtering spintronics were discussed.

\bibliographystyle{apsrev4-2}

\begin{thebibliography}{31}%
\makeatletter
\providecommand \@ifxundefined [1]{%
 \@ifx{#1\undefined}
}%
\providecommand \@ifnum [1]{%
 \ifnum #1\expandafter \@firstoftwo
 \else \expandafter \@secondoftwo
 \fi
}%
\providecommand \@ifx [1]{%
 \ifx #1\expandafter \@firstoftwo
 \else \expandafter \@secondoftwo
 \fi
}%
\providecommand \natexlab [1]{#1}%
\providecommand \enquote  [1]{``#1''}%
\providecommand \bibnamefont  [1]{#1}%
\providecommand \bibfnamefont [1]{#1}%
\providecommand \citenamefont [1]{#1}%
\providecommand \href@noop [0]{\@secondoftwo}%
\providecommand \href [0]{\begingroup \@sanitize@url \@href}%
\providecommand \@href[1]{\@@startlink{#1}\@@href}%
\providecommand \@@href[1]{\endgroup#1\@@endlink}%
\providecommand \@sanitize@url [0]{\catcode `\\12\catcode `\$12\catcode
  `\&12\catcode `\#12\catcode `\^12\catcode `\_12\catcode `\%12\relax}%
\providecommand \@@startlink[1]{}%
\providecommand \@@endlink[0]{}%
\providecommand \url  [0]{\begingroup\@sanitize@url \@url }%
\providecommand \@url [1]{\endgroup\@href {#1}{\urlprefix }}%
\providecommand \urlprefix  [0]{URL }%
\providecommand \Eprint [0]{\href }%
\providecommand \doibase [0]{https://doi.org/}%
\providecommand \selectlanguage [0]{\@gobble}%
\providecommand \bibinfo  [0]{\@secondoftwo}%
\providecommand \bibfield  [0]{\@secondoftwo}%
\providecommand \translation [1]{[#1]}%
\providecommand \BibitemOpen [0]{}%
\providecommand \bibitemStop [0]{}%
\providecommand \bibitemNoStop [0]{.\EOS\space}%
\providecommand \EOS [0]{\spacefactor3000\relax}%
\providecommand \BibitemShut  [1]{\csname bibitem#1\endcsname}%
\let\auto@bib@innerbib\@empty
%
\bibitem{C1987}Coey, J. M. D. Noncollinear spin structures. Can. J. Phys. 65, 1210–1232 (1987).
%
\bibitem{JMW2016} Jungwirth, T., Marti, X., Wadley, P. and Wunderlich, J. Antiferromagnetic spintronics. Nature Nanotechnol. 11, 231–241 (2016)
%
 \bibitem{RSL2000} M. P. Recher, E. V. Sukhorukov, and D. Loss, {Phys. Rev. Lett.} {\bf 85}, 1962 (2000)
 \bibitem{SSG2002} N. Sergueev,  Qing-feng Sun,  Hong Guo,  B. G. Wang,  and Jian Wang, {Phys. Rev. B} {\bf 65}, 165303 (2002)
 \bibitem{FHE2003} J. Fransson, E. Holmstro\"m, O. Eriksson, and I. Sandalov {Phys. Rev. B} {\bf 67}, 205310 (2003) 
 \bibitem{PB2006} M. Pustilnik and L. Borda, {Phys. Rev. B} {\bf 73}, 201301R (2006)
  
 \bibitem{SCA2006} R. S\'anchez, E. Cota, R. Aguado,  and G. Platero, {Phys. Rev. B} {\bf 74}, 035326 (2006)
%
\bibitem{WCA2023} Winkler, R., Ciria, M., Ahmad, M., Plank, H., Marcuello, C.
{Nanomaterials}  {\bf 13}, 2585 (2023) 
https://doi.org/10.3390/nano13182585 
%
\bibitem{WB2001}
R Wiesendanger, M Bode,
{Solid State Communications},
{\bf 119}, 341 (2001)
https://doi.org/10.1016/S0038-1098(01)00103-X.
%
\bibitem{GAG2017} Gross, I., Akhtar, W., Garcia, V. et al. ,
{Nature} {\bf 549}, 252 (2017)
%
\bibitem{GSP2023} Ge, Z., Slizovskiy, S., Polizogopoulos, P. et al., 
{Nat. Nanotechnol. } {\bf 18}, 250 (2023)
%
\bibitem [{\citenamefont {Averin}\ and\ \citenamefont
  {Likharev}(1986)}]{AL1986}%
  \BibitemOpen
  \bibfield  {author} {\bibinfo {author} {\bibfnamefont {D.~V.}\ \bibnamefont
  {Averin}}\ and\ \bibinfo {author} {\bibfnamefont {K.~K.}\ \bibnamefont
  {Likharev}},\ }\href {https://doi.org/10.1007/BF00683469} {\bibfield
  {journal} {\bibinfo  {journal} {Journal of Low Temperature Physics}\ }\textbf
  {\bibinfo {volume} {62}},\ \bibinfo {pages} {345} (\bibinfo {year}
  {1986})}\BibitemShut {NoStop}%
\bibitem [{\citenamefont {Fulton}\ and\ \citenamefont {Dolan}(1987)}]{FD1987}%
  \BibitemOpen
  \bibfield  {author} {\bibinfo {author} {\bibfnamefont {T.~A.}\ \bibnamefont
  {Fulton}}\ and\ \bibinfo {author} {\bibfnamefont {G.~J.}\ \bibnamefont
  {Dolan}},\ }\href {https://doi.org/10.1103/PhysRevLett.59.109} {\bibfield
  {journal} {\bibinfo  {journal} {Phys. Rev. Lett.}\ }\textbf {\bibinfo
  {volume} {59}},\ \bibinfo {pages} {109} (\bibinfo {year} {1987})}\BibitemShut
  {NoStop}%
\bibitem [{\citenamefont {Lafarge}\ \emph {et~al.}(1991)\citenamefont
  {Lafarge}, \citenamefont {Pothier}, \citenamefont {Williams}, \citenamefont
  {Esteve}, \citenamefont {Urbina},\ and\ \citenamefont
  {Devoret}}]{LPWEUD1991}%
  \BibitemOpen
  \bibfield  {author} {\bibinfo {author} {\bibfnamefont {P.}~\bibnamefont
  {Lafarge}}, \bibinfo {author} {\bibfnamefont {H.}~\bibnamefont {Pothier}},
  \bibinfo {author} {\bibfnamefont {E.~R.}\ \bibnamefont {Williams}}, \bibinfo
  {author} {\bibfnamefont {D.}~\bibnamefont {Esteve}}, \bibinfo {author}
  {\bibfnamefont {C.}~\bibnamefont {Urbina}},\ and\ \bibinfo {author}
  {\bibfnamefont {M.~H.}\ \bibnamefont {Devoret}},\ }\href
  {https://doi.org/10.1007/BF01307627} {\bibfield  {journal} {\bibinfo
  {journal} {Zeitschrift f{\"u}r Physik B Condensed Matter}\ }\textbf {\bibinfo
  {volume} {85}},\ \bibinfo {pages} {327} (\bibinfo {year} {1991})}\BibitemShut
  {NoStop}%
\bibitem [{\citenamefont {Kastner}(1992)}]{Kastner1992}%
  \BibitemOpen
  \bibfield  {author} {\bibinfo {author} {\bibfnamefont {M.~A.}\ \bibnamefont
  {Kastner}},\ }\href {https://doi.org/10.1103/RevModPhys.64.849} {\bibfield
  {journal} {\bibinfo  {journal} {Rev. Mod. Phys.}\ }\textbf {\bibinfo {volume}
  {64}},\ \bibinfo {pages} {849} (\bibinfo {year} {1992})}\BibitemShut
  {NoStop}%
\bibitem [{\citenamefont {Hewson}(1993)}]{Hewson1993book}%
  \BibitemOpen
  \bibfield  {author} {\bibinfo {author} {\bibfnamefont {A.~C.}\ \bibnamefont
  {Hewson}},\ }\href@noop {} {\emph {\bibinfo {title} {The Kondo Problem to
  Heavy Fermions}}}\ (\bibinfo  {publisher} {Cambridge University Press},\
  \bibinfo {address} {Cambridge, England},\ \bibinfo {year} {1993})\BibitemShut
  {NoStop}%
\bibitem [{\citenamefont {Goldhaber-Gordon}\ \emph {et~al.}(1998)\citenamefont
  {Goldhaber-Gordon}, \citenamefont {Shtrikman}, \citenamefont {Mahalu},
  \citenamefont {Abusch-Magder}, \citenamefont {Meirav},\ and\ \citenamefont
  {Kastner}}]{GSM1998}%
  \BibitemOpen
  \bibfield  {author} {\bibinfo {author} {\bibfnamefont {D.}~\bibnamefont
  {Goldhaber-Gordon}}, \bibinfo {author} {\bibfnamefont {H.}~\bibnamefont
  {Shtrikman}}, \bibinfo {author} {\bibfnamefont {D.}~\bibnamefont {Mahalu}},
  \bibinfo {author} {\bibfnamefont {D.}~\bibnamefont {Abusch-Magder}}, \bibinfo
  {author} {\bibfnamefont {U.}~\bibnamefont {Meirav}},\ and\ \bibinfo {author}
  {\bibfnamefont {M.~A.}\ \bibnamefont {Kastner}},\ }\href
  {https://doi.org/10.1038/34373} {\bibfield  {journal} {\bibinfo  {journal}
  {Nature}\ }\textbf {\bibinfo {volume} {391}},\ \bibinfo {pages} {156}
  (\bibinfo {year} {1998})}\BibitemShut {NoStop}%
\bibitem [{\citenamefont {Klein}\ \emph {et~al.}(2007)\citenamefont {Klein},
  \citenamefont {Savage},\ and\ \citenamefont {Eriksson}}]{KSE2007}%
  \BibitemOpen
  \bibfield  {author} {\bibinfo {author} {\bibfnamefont {L.~J.}\ \bibnamefont
  {Klein}}, \bibinfo {author} {\bibfnamefont {D.~E.}\ \bibnamefont {Savage}},\
  and\ \bibinfo {author} {\bibfnamefont {M.~A.}\ \bibnamefont {Eriksson}},\
  }\href {https://doi.org/10.1063/1.2431760} {\bibfield  {journal} {\bibinfo
  {journal} {Appl. Phys. Lett.}\ }\textbf {\bibinfo {volume} {90}},\ \bibinfo
  {pages} {033103} (\bibinfo {year} {2007})}\BibitemShut {NoStop}%
\bibitem [{\citenamefont {Liang}\ \emph {et~al.}(2002)\citenamefont {Liang},
  \citenamefont {Shores}, \citenamefont {Bockrath}, \citenamefont {Long},\ and\
  \citenamefont {Park}}]{LSB2002}%
  \BibitemOpen
  \bibfield  {author} {\bibinfo {author} {\bibfnamefont {W.}~\bibnamefont
  {Liang}}, \bibinfo {author} {\bibfnamefont {M.}~\bibnamefont {Shores}},
  \bibinfo {author} {\bibfnamefont {M.}~\bibnamefont {Bockrath}}, \bibinfo
  {author} {\bibfnamefont {J.}~\bibnamefont {Long}},\ and\ \bibinfo {author}
  {\bibfnamefont {H.}~\bibnamefont {Park}},\ }\href
  {https://doi.org/10.1038/nature00790} {\bibfield  {journal} {\bibinfo
  {journal} {Nature}\ }\textbf {\bibinfo {volume} {417}},\ \bibinfo {pages}
  {725} (\bibinfo {year} {2002})}\BibitemShut {NoStop}%
\bibitem [{\citenamefont {Nygard}\ \emph {et~al.}(2000)\citenamefont {Nygard},
  \citenamefont {Cobden},\ and\ \citenamefont {Lindelof}}]{NCL2000}%
  \BibitemOpen
  \bibfield  {author} {\bibinfo {author} {\bibfnamefont {J.}~\bibnamefont
  {Nygard}}, \bibinfo {author} {\bibfnamefont {D.~H.}\ \bibnamefont {Cobden}},\
  and\ \bibinfo {author} {\bibfnamefont {P.~E.}\ \bibnamefont {Lindelof}},\
  }\href {https://doi.org/10.1038/35042545} {\bibfield  {journal} {\bibinfo
  {journal} {Nature}\ }\textbf {\bibinfo {volume} {408}},\ \bibinfo {pages}
  {342} (\bibinfo {year} {2000})}\BibitemShut {NoStop}%
\bibitem [{\citenamefont {van~der Wiel}\ \emph {et~al.}(2000)\citenamefont
  {van~der Wiel}, \citenamefont {De~Franceschi}, \citenamefont {Fujisawa},
  \citenamefont {Elzerman}, \citenamefont {Tarucha},\ and\ \citenamefont
  {Kouwenhoven}}]{WFF2000}%
  \BibitemOpen
  \bibfield  {author} {\bibinfo {author} {\bibfnamefont {W.}~\bibnamefont
  {van~der Wiel}}, \bibinfo {author} {\bibfnamefont {S.}~\bibnamefont
  {De~Franceschi}}, \bibinfo {author} {\bibfnamefont {T.}~\bibnamefont
  {Fujisawa}}, \bibinfo {author} {\bibfnamefont {J.~M.}\ \bibnamefont
  {Elzerman}}, \bibinfo {author} {\bibfnamefont {S.}~\bibnamefont {Tarucha}},\
  and\ \bibinfo {author} {\bibfnamefont {L.~P.}\ \bibnamefont {Kouwenhoven}},\
  }\href {https://doi.org/10.1126/science.289.5487.2105} {\bibfield  {journal}
  {\bibinfo  {journal} {Science}\ }\textbf {\bibinfo {volume} {289}},\ \bibinfo
  {pages} {2105} (\bibinfo {year} {2000})}\BibitemShut {NoStop}%
\bibitem [{\citenamefont {Klochan}\ \emph {et~al.}(2013)\citenamefont
  {Klochan}, \citenamefont {Micolich}, \citenamefont {Hamilton}, \citenamefont
  {Reuter}, \citenamefont {Wieck}, \citenamefont {Reininghaus}, \citenamefont
  {Pletyukhov},\ and\ \citenamefont {Schoeller}}]{KMH2013}%
  \BibitemOpen
  \bibfield  {author} {\bibinfo {author} {\bibfnamefont {O.}~\bibnamefont
  {Klochan}}, \bibinfo {author} {\bibfnamefont {A.~P.}\ \bibnamefont
  {Micolich}}, \bibinfo {author} {\bibfnamefont {A.~R.}\ \bibnamefont
  {Hamilton}}, \bibinfo {author} {\bibfnamefont {D.}~\bibnamefont {Reuter}},
  \bibinfo {author} {\bibfnamefont {A.~D.}\ \bibnamefont {Wieck}}, \bibinfo
  {author} {\bibfnamefont {F.}~\bibnamefont {Reininghaus}}, \bibinfo {author}
  {\bibfnamefont {M.}~\bibnamefont {Pletyukhov}},\ and\ \bibinfo {author}
  {\bibfnamefont {H.}~\bibnamefont {Schoeller}},\ }\href
  {https://doi.org/10.1103/PhysRevB.87.201104} {\bibfield  {journal} {\bibinfo
  {journal} {Phys. Rev. B}\ }\textbf {\bibinfo {volume} {87}},\ \bibinfo
  {pages} {201104} (\bibinfo {year} {2013})}\BibitemShut {NoStop}%
\bibitem [{\citenamefont {Daroca}\ \emph {et~al.}(2018)\citenamefont {Daroca},
  \citenamefont {Roura-Bas},\ and\ \citenamefont {Aligia}}]{DRA2018}%
  \BibitemOpen
  \bibfield  {author} {\bibinfo {author} {\bibfnamefont {D.~P.}\ \bibnamefont
  {Daroca}}, \bibinfo {author} {\bibfnamefont {P.}~\bibnamefont {Roura-Bas}},\
  and\ \bibinfo {author} {\bibfnamefont {A.~A.}\ \bibnamefont {Aligia}},\
  }\href {https://doi.org/10.1103/PhysRevB.98.245406} {\bibfield  {journal}
  {\bibinfo  {journal} {Phys. Rev. B}\ }\textbf {\bibinfo {volume} {98}},\
  \bibinfo {pages} {245406} (\bibinfo {year} {2018})}\BibitemShut {NoStop}%
\bibitem [{\citenamefont {Mathew}\ and\ \citenamefont {Fang}(2018)}]{MF2018}%
  \BibitemOpen
  \bibfield  {author} {\bibinfo {author} {\bibfnamefont {P.~T.}\ \bibnamefont
  {Mathew}}\ and\ \bibinfo {author} {\bibfnamefont {F.}~\bibnamefont {Fang}},\
  }\href {https://doi.org/https://doi.org/10.1016/j.eng.2018.11.001} {\bibfield
   {journal} {\bibinfo  {journal} {Engineering}\ }\textbf {\bibinfo {volume}
  {4}},\ \bibinfo {pages} {760 } (\bibinfo {year} {2018})}\BibitemShut
  {NoStop}%
\bibitem [{\citenamefont {Limburg}\ \emph {et~al.}(2019)\citenamefont
  {Limburg}, \citenamefont {Thomas}, \citenamefont {Sowa}, \citenamefont
  {Willick}, \citenamefont {Baugh}, \citenamefont {Gauger}, \citenamefont
  {Briggs}, \citenamefont {Mol},\ and\ \citenamefont {Anderson}}]{LTS2019}%
  \BibitemOpen
  \bibfield  {author} {\bibinfo {author} {\bibfnamefont {B.}~\bibnamefont
  {Limburg}}, \bibinfo {author} {\bibfnamefont {J.~O.}\ \bibnamefont {Thomas}},
  \bibinfo {author} {\bibfnamefont {J.~K.}\ \bibnamefont {Sowa}}, \bibinfo
  {author} {\bibfnamefont {K.}~\bibnamefont {Willick}}, \bibinfo {author}
  {\bibfnamefont {J.}~\bibnamefont {Baugh}}, \bibinfo {author} {\bibfnamefont
  {E.~M.}\ \bibnamefont {Gauger}}, \bibinfo {author} {\bibfnamefont {G.~A.~D.}\
  \bibnamefont {Briggs}}, \bibinfo {author} {\bibfnamefont {J.~A.}\
  \bibnamefont {Mol}},\ and\ \bibinfo {author} {\bibfnamefont {H.~L.}\
  \bibnamefont {Anderson}},\ }\href {https://doi.org/10.1039/C9NR03754C}
  {\bibfield  {journal} {\bibinfo  {journal} {Nanoscale}\ }\textbf {\bibinfo
  {volume} {11}},\ \bibinfo {pages} {14820} (\bibinfo {year}
  {2019})}\BibitemShut {NoStop}%
%
\bibitem [{\citenamefont {Yu}\ and\ \citenamefont {Natelson}(2004)}]{YN2004}%
  \BibitemOpen
  \bibfield  {author} {\bibinfo {author} {\bibfnamefont {L.~H.}\ \bibnamefont
  {Yu}}\ and\ \bibinfo {author} {\bibfnamefont {D.}~\bibnamefont {Natelson}},\
  }\href {https://doi.org/10.1021/nl034893f} {\bibfield  {journal} {\bibinfo
  {journal} {Nano Letters}\ }\textbf {\bibinfo {volume} {4}},\ \bibinfo {pages}
  {79} (\bibinfo {year} {2004})}\BibitemShut {NoStop}%
%
\bibitem [{\citenamefont {Wahl}\ \emph {et~al.}(2004)\citenamefont {Wahl},
  \citenamefont {Diekh\"oner}, \citenamefont {Schneider}, \citenamefont
  {Vitali}, \citenamefont {Wittich},\ and\ \citenamefont {Kern}}]{WDS2004}%
  \BibitemOpen
  \bibfield  {author} {\bibinfo {author} {\bibfnamefont {P.}~\bibnamefont
  {Wahl}}, \bibinfo {author} {\bibfnamefont {L.}~\bibnamefont {Diekh\"oner}},
  \bibinfo {author} {\bibfnamefont {M.~A.}\ \bibnamefont {Schneider}}, \bibinfo
  {author} {\bibfnamefont {L.}~\bibnamefont {Vitali}}, \bibinfo {author}
  {\bibfnamefont {G.}~\bibnamefont {Wittich}},\ and\ \bibinfo {author}
  {\bibfnamefont {K.}~\bibnamefont {Kern}},\ }\href
  {https://doi.org/10.1103/PhysRevLett.93.176603} {\bibfield  {journal}
  {\bibinfo  {journal} {Phys. Rev. Lett.}\ }\textbf {\bibinfo {volume} {93}},\
  \bibinfo {pages} {176603} (\bibinfo {year} {2004})}\BibitemShut {NoStop}%
\bibitem [{\citenamefont {Tie}\ \emph {et~al.}(2019)\citenamefont {Tie},
  \citenamefont {Gravelsins}, \citenamefont {Niewczas},\ and\ \citenamefont
  {Dhirani}}]{TGN2019}%
  \BibitemOpen
  \bibfield  {author} {\bibinfo {author} {\bibfnamefont {M.}~\bibnamefont
  {Tie}}, \bibinfo {author} {\bibfnamefont {S.}~\bibnamefont {Gravelsins}},
  \bibinfo {author} {\bibfnamefont {M.}~\bibnamefont {Niewczas}},\ and\
  \bibinfo {author} {\bibfnamefont {A.}~\bibnamefont {Dhirani}},\ }\href
  {https://doi.org/10.1039/c8nr09280j} {\bibfield  {journal} {\bibinfo
  {journal} {Nanoscale}\ }\textbf {\bibinfo {volume} {11}},\ \bibinfo {pages}
  {5395} (\bibinfo {year} {2019})}\BibitemShut {NoStop}%
%
\bibitem{SG2013} S. Smirnov and M. Grifoni,
{New Journal of Physics} {\bf 15}, 073047 (2022)
%
\bibitem{DAM2019} B. Danu, F. Assaad, and F. Mila,
{Phys. Rev. Lett.} {\bf 123}, 176601 (2019)
%
\bibitem{ZKH2013} Zhang, Yh., Kahle, S., Herden, T. et al., 
{Nat Commun } {\bf 4}, 2110 (2013)
%
\bibitem{HCV2022} C. Hsu, T. A. Costi, D. Vogel, C. Wegeberg, M. Mayor, H.S.J. van der Zant, and P. Gehring,
{Phys. Rev. Lett.} {\bf 128}, 147701 (2022)
%
\bibitem [{\citenamefont {Chan}\ \emph {et~al.}(2002)\citenamefont {Chan},
  \citenamefont {Westervelt}, \citenamefont {Maranowski},\ and\ \citenamefont
  {Gossard}}]{CWM2002}%
  \BibitemOpen
  \bibfield  {author} {\bibinfo {author} {\bibfnamefont {I.~H.}\ \bibnamefont
  {Chan}}, \bibinfo {author} {\bibfnamefont {R.~M.}\ \bibnamefont
  {Westervelt}}, \bibinfo {author} {\bibfnamefont {K.~D.}\ \bibnamefont
  {Maranowski}},\ and\ \bibinfo {author} {\bibfnamefont {A.~C.}\ \bibnamefont
  {Gossard}},\ }\href {https://doi.org/10.1063/1.1456552} {\bibfield  {journal}
  {\bibinfo  {journal} {Appl. Phys. Lett.}\ }\textbf {\bibinfo {volume} {80}},\
  \bibinfo {pages} {1818} (\bibinfo {year} {2002})}\BibitemShut {NoStop}%
\bibitem [{\citenamefont {H\"ubel}\ \emph {et~al.}(2007)\citenamefont
  {H\"ubel}, \citenamefont {Weis}, \citenamefont {Dietsche},\ and\
  \citenamefont {Klitzing}}]{HWDK2007}%
  \BibitemOpen
  \bibfield  {author} {\bibinfo {author} {\bibfnamefont {A.}~\bibnamefont
  {H\"ubel}}, \bibinfo {author} {\bibfnamefont {J.}~\bibnamefont {Weis}},
  \bibinfo {author} {\bibfnamefont {W.}~\bibnamefont {Dietsche}},\ and\
  \bibinfo {author} {\bibfnamefont {K.~v.}\ \bibnamefont {Klitzing}},\ }\href
  {https://doi.org/10.1063/1.2778542} {\bibfield  {journal} {\bibinfo
  {journal} {Appl. Phys. Lett.}\ }\textbf {\bibinfo {volume} {91}},\ \bibinfo
  {pages} {102101} (\bibinfo {year} {2007})}\BibitemShut {NoStop}%
\bibitem [{\citenamefont {Fringes}\ \emph {et~al.}(2012)\citenamefont
  {Fringes}, \citenamefont {Volk}, \citenamefont {Terr\'es}, \citenamefont
  {Dauber}, \citenamefont {Engels}, \citenamefont {Trellenkamp},\ and\
  \citenamefont {Stampfer}}]{FVT2012}%
  \BibitemOpen
  \bibfield  {author} {\bibinfo {author} {\bibfnamefont {S.}~\bibnamefont
  {Fringes}}, \bibinfo {author} {\bibfnamefont {C.}~\bibnamefont {Volk}},
  \bibinfo {author} {\bibfnamefont {B.}~\bibnamefont {Terr\'es}}, \bibinfo
  {author} {\bibfnamefont {J.}~\bibnamefont {Dauber}}, \bibinfo {author}
  {\bibfnamefont {S.}~\bibnamefont {Engels}}, \bibinfo {author} {\bibfnamefont
  {S.}~\bibnamefont {Trellenkamp}},\ and\ \bibinfo {author} {\bibfnamefont
  {C.}~\bibnamefont {Stampfer}},\ }\href
  {https://doi.org/10.1002/pssc.201100340} {\bibfield  {journal} {\bibinfo
  {journal} {Phys. Status Solidi C}\ }\textbf {\bibinfo {volume} {9}},\
  \bibinfo {pages} {169–174} (\bibinfo {year} {2012})}\BibitemShut {NoStop}%
\bibitem [{\citenamefont {Anderson}(1961)}]{A1961}%
  \BibitemOpen
  \bibfield  {author} {\bibinfo {author} {\bibfnamefont {P.~W.}\ \bibnamefont
  {Anderson}},\ }\href {https://doi.org/10.1103/PhysRev.124.41} {\bibfield
  {journal} {\bibinfo  {journal} {Phys. Rev.}\ }\textbf {\bibinfo {volume}
  {124}},\ \bibinfo {pages} {41} (\bibinfo {year} {1961})}\BibitemShut
  {NoStop}%
\bibitem [{\citenamefont {Maier}\ \emph {et~al.}(2000)\citenamefont {Maier},
  \citenamefont {Jarrell}, \citenamefont {Pruschke},\ and\ \citenamefont
  {Keller}}]{MJP2000}%
  \BibitemOpen
  \bibfield  {author} {\bibinfo {author} {\bibfnamefont {T.}~\bibnamefont
  {Maier}}, \bibinfo {author} {\bibfnamefont {M.}~\bibnamefont {Jarrell}},
  \bibinfo {author} {\bibfnamefont {T.}~\bibnamefont {Pruschke}},\ and\
  \bibinfo {author} {\bibfnamefont {J.}~\bibnamefont {Keller}},\ }\href
  {https://doi.org/10.1007/s100510050077} {\bibfield  {journal} {\bibinfo
  {journal} {Euro. Phys. J. B}\ }\textbf {\bibinfo {volume} {13}},\ \bibinfo
  {pages} {613} (\bibinfo {year} {2000})}\BibitemShut {NoStop}%
\bibitem [{\citenamefont {Otsuki}\ and\ \citenamefont
  {Kuramoto}(2006)}]{OK2006}%
  \BibitemOpen
  \bibfield  {author} {\bibinfo {author} {\bibfnamefont {J.}~\bibnamefont
  {Otsuki}}\ and\ \bibinfo {author} {\bibfnamefont {Y.}~\bibnamefont
  {Kuramoto}},\ }\href {https://doi.org/10.1143/JPSJ.75.064707} {\bibfield
  {journal} {\bibinfo  {journal} {J. Phys. Soc. Jpn}\ }\textbf {\bibinfo
  {volume} {75}},\ \bibinfo {pages} {064707} (\bibinfo {year}
  {2006})}\BibitemShut {NoStop}%
%
\bibitem [{\citenamefont {Zamani}\ \emph {et~al.}(2013)\citenamefont {Zamani},
  \citenamefont {Chowdhury}, \citenamefont {Ribeiro}, \citenamefont
  {Ingersent},\ and\ \citenamefont {Kirchner}}]{ZCR2013}%
  \BibitemOpen
  \bibfield  {author} {\bibinfo {author} {\bibfnamefont {F.}~\bibnamefont
  {Zamani}}, \bibinfo {author} {\bibfnamefont {T.}~\bibnamefont {Chowdhury}},
  \bibinfo {author} {\bibfnamefont {P.}~\bibnamefont {Ribeiro}}, \bibinfo
  {author} {\bibfnamefont {K.}~\bibnamefont {Ingersent}},\ and\ \bibinfo
  {author} {\bibfnamefont {S.}~\bibnamefont {Kirchner}},\ }\href
  {https://doi.org/10.1002/pssb.201200928} {\bibfield  {journal} {\bibinfo
  {journal} {Phys. Status Solidi B}\ }\textbf {\bibinfo {volume} {250}},\
  \bibinfo {pages} {547} (\bibinfo {year} {2013})}\BibitemShut {NoStop}%
%
\bibitem{YWD2011} S. Yang, X. Wang, and S. Das Sarma, 
{Phys. Rev. B} {\bf 83}, 161301(R) (2011)
%
  \bibitem [{\citenamefont {Jauho}\ \emph {et~al.}(1994)\citenamefont {Jauho},
  \citenamefont {Wingreen},\ and\ \citenamefont
  {Meir}}]{JauhoWingreenMeir1994}%
  \BibitemOpen
  \bibfield  {author} {\bibinfo {author} {\bibfnamefont {A.-P.}\ \bibnamefont
  {Jauho}}, \bibinfo {author} {\bibfnamefont {N.~S.}\ \bibnamefont
  {Wingreen}},\ and\ \bibinfo {author} {\bibfnamefont {Y.}~\bibnamefont
  {Meir}},\ }\href {https://doi.org/10.1103/PhysRevB.50.5528} {\bibfield
  {journal} {\bibinfo  {journal} {Phys. Rev. B}\ }\textbf {\bibinfo {volume}
  {50}},\ \bibinfo {pages} {5528} (\bibinfo {year} {1994})}\BibitemShut
  {NoStop}%
\bibitem [{\citenamefont {Kim}\ and\ \citenamefont
  {Hershfield}(2003)}]{KH2003}%
  \BibitemOpen
  \bibfield  {author} {\bibinfo {author} {\bibfnamefont {T.-S.}\ \bibnamefont
  {Kim}}\ and\ \bibinfo {author} {\bibfnamefont {S.}~\bibnamefont
  {Hershfield}},\ }\href {https://doi.org/10.1103/PhysRevB.67.165313}
  {\bibfield  {journal} {\bibinfo  {journal} {Phys. Rev. B}\ }\textbf {\bibinfo
  {volume} {67}},\ \bibinfo {pages} {165313} (\bibinfo {year}
  {2003})}\BibitemShut {NoStop}%
\bibitem [{\citenamefont {Wingreen}\ \emph {et~al.}(1993)\citenamefont
  {Wingreen}, \citenamefont {Jauho},\ and\ \citenamefont {Meir}}]{WJM1993}%
  \BibitemOpen
  \bibfield  {author} {\bibinfo {author} {\bibfnamefont {N.~S.}\ \bibnamefont
  {Wingreen}}, \bibinfo {author} {\bibfnamefont {A.~P.}\ \bibnamefont
  {Jauho}},\ and\ \bibinfo {author} {\bibfnamefont {Y.}~\bibnamefont {Meir}},\
  }\href {https://doi.org/10.1103/PhysRevB.48.8487} {\bibfield  {journal}
  {\bibinfo  {journal} {Phys. Rev. B}\ }\textbf {\bibinfo {volume} {48}},\
  \bibinfo {pages} {8487} (\bibinfo {year} {1993})}\BibitemShut {NoStop}%
 %
\end{thebibliography}

%

\end{document}